# Visualization and manipulation of four-leaf clover-shaped electronic state in cuprate

Zechao Wang[1, †, *], Fengyu Yao[1, †], Yuchen Huo[1], Zhongxu Wei[4], Zhiyuan Song[1], Mingqiang Ren[1,2], Ziyuan Cheng[1], Jinfeng Jia[1,2,5], Yu-Jie Sun[1,2*], Qi-Kun Xue[1,2,3 *]

1. Department of Physics, Southern University of Science and Technology, Shenzhen, P.R. China.
2. Quantum Science Center of Guangdong-HongKong-Macao Greater Bay Area, Shenzhen, P. R. China.
3. State Key Laboratory of Low-Dimensional Quantum Physics, Department of Physics, Tsinghua University, Beijing, P. R. China
4. Beijing National Laboratory for Condensed Matter Physics, Institute of Physics, Chinese Academy of Sciences, Beijing, P. R. China.
5. Department of Physics and Astronomy, Shanghai Jiao Tong University, Shanghai, P. R. China.

† These authors contributed equally to this work.

* To whom correspondence should be addressed: wangzc@sustech.edu.cn; sunyj@sustech.edu.cn; xueqk@sustech.edu.cn.

**High-$T_c$ superconductivity in cuprates arises from carrier doping of an antiferromagnetic Mott insulator. Associated with these changes are spectral-weight transfers from the high-energy to low-energy, giving rise to a variety of intriguing electronic phenomena. In this study, for the first time, we discovered a $2a_0$ sized four-leaf clover-shaped (FLC) electronic state at low-energy, accompanied with the emergence of a characteristic "kink" around 16meV. With increasing doping, the number of FLC pattern decreases and ultimately vanishes in the overdoped region. Remarkably, we achieved real-time electric-field manipulation of this FLC state, through innovative in-situ scanning tunneling microscopy probe. This novel FLC state may not only redefine our understanding of precursor states of pairing, but also reveals its crucial role as a tunable electronic phase in high-$T_c$ superconductors.**

Owing to strong correlation effects, the parent phase of cuprates is an antiferromagnetic (AFM) charge-transfer-type insulator (*1,2,3*). Upon doping holes into this parent phase, the long-range AFM order is gradually suppressed, leading to a transfer of the spectral weight from the upper Hubbard band to low energies (*4-7*)(as shown in Fig.1D). This spectral-weight transfers along with the emergence of a broadened in-gap state (*8*), and accompanied by the opening of a gap around the Fermi level and with further doping, superconductivity emerges. Understanding the behavior of doped holes and the associated electronic order within the AFM Mott insulator background is crucial for elucidating the mechanism of high-$T_c$ superconductivity (*9*).

Experimental investigations on hole-doped copper oxides, particularly those with a large charge-transfer gap, have predominantly focused on the formation of charge order. In early studies of La- and Y-based superconductors, unidirectional fluctuating stripe phases were observed (*10,11,12*), while similar stripe and checkerboard electronic patterns were later discovered in Bi- and Cl-based superconductors within the pseudogap (PG) region (*13-26*). Various explanations have been proposed to explain these charge-ordered states, which span a wide doping and energy range in copper oxides. This includes analogies to electronic quantum liquid-phase crystal that break certain symmetries, such as nematic or smectic phases (*27*), instabilities arising from Fermi surface nesting (*23*), or disorder-pinned electronic-cluster-glass (ECG) (*16*).

However, experimental evidence indicates that the spectral-weight transfers induced intertwined order is closely associated with local Cooper pair formation (*28,29*) or other forms of carriers, such as the possible Zhang-Rice singlet (ZRS) (*30*), as well as its potential interplay with global superconductivity, remains insufficiently understood. To address these questions, we utilized high-resolution scanning tunneling microscopy (STM) to investigate the low-energy local electronic behavior of $Bi_2Sr_2CaCu_2O_{8+\delta}$ (Bi-2212) at various doping levels, from insulating to overdoped superconducting phase, as illustrate in the phase diagram in Fig.1A.

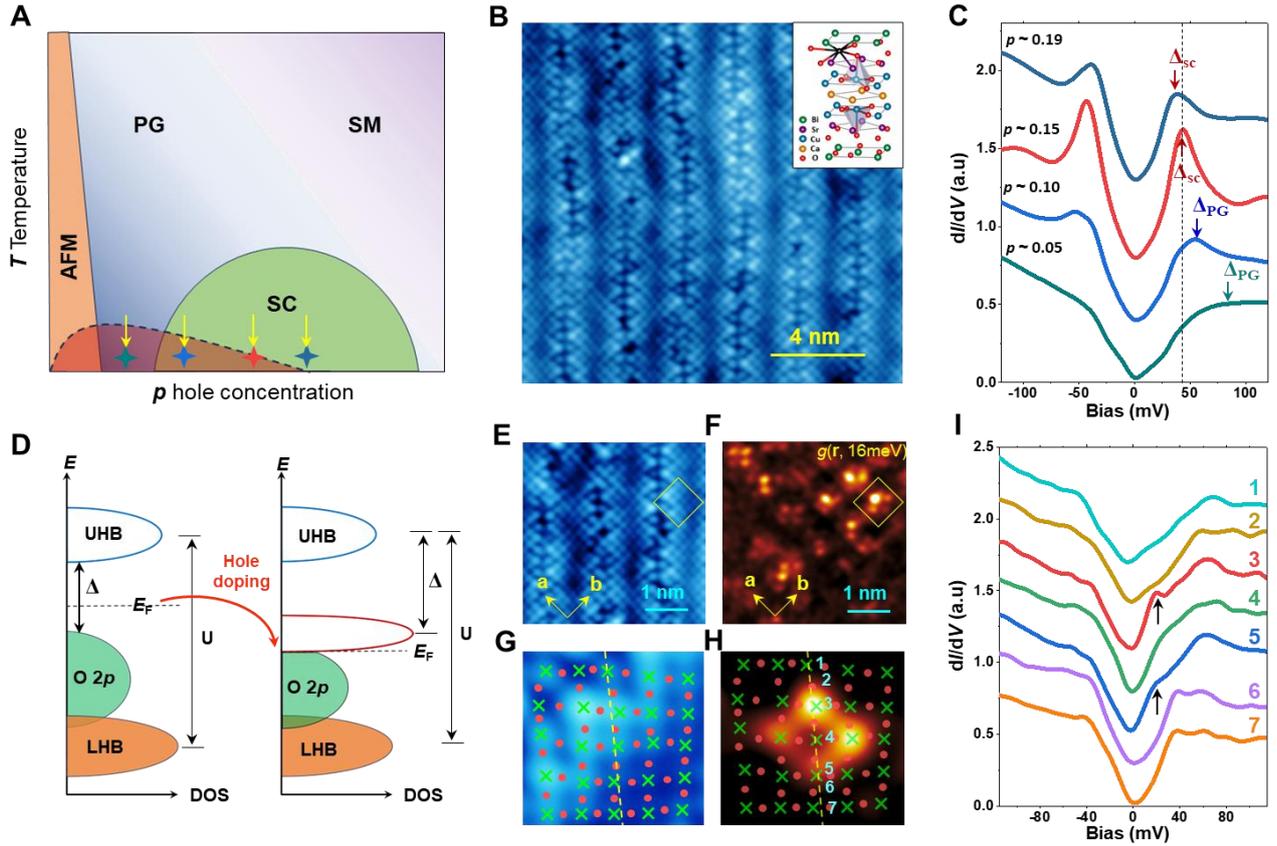

**Fig.1. Phase diagram, band structure and electronic structure of a FLC state.** (**A**) Schematic phase diagram of hole-doped cuprates. The regions of AFM insulator, superconductivity (SC), and pseudogap (PG) region are indicated. Arrows indicate the hole densities of the four samples studied in this work. (**B**) Atomic-scale high-resolution topography of the BiO cleaved surface and the inset shows the the crystal structure model of Bi-2212 with near one dopant oxygen. The dopant oxygen is indicated by black atom. (**C**) The averaged spectra of the four samples after high-temperature annealing corresponds to an estimated hole concentration of $p\sim0.05$, $p\sim0.10$, $p\sim0.15$ and $p\sim0.19$, respectively. (**D**) Schematic band structure in the $CuO_2$ plane and the evolution of hole doping effects upon it. (**E**) Topographic image of a $p\sim0.05$ Bi-2212 sample. (**F**) The density of state (DOS) mapping at 16 mV in the same field of view as (E). (**G** and **H**) Zoomed-in topography and DOS map of the yellow boxes areas as shown in (E and F), respectively. The zoomed-in DOS map displaying a FLC electron pattern centered around a Cu site. The Cu atom and oxygen atoms are marked by green cross and red dot, respectively. (**I**) A series of d$I$/d$V$ spectra taken along the yellow dotted line in (G and H). The spectral weight of low energy state is indicated by black arrows. All data acquired at $T = 4.2$ K.

## Discovery of a $2a_0$ sized FLC electronic state at low-energy.

The layered structure of $Bi_2Sr_2CaCu_2O_{8+\delta}$ (Bi-2212) is an excellent material for investigating the microscopic pairing mechanism in high-$T_c$ superconductors. Due to the weak van der Waals bonding between adjacent Bi-O planes in Bi-family cuprates, the electronic structure of the cleaved surface has been shown to reflect the bulk properties (*31,32*). Fig. 1B shows the typical topography of the heavily underdoped sample ($p$~0.05), where the supermodulation along the [110] direction of the exposed BiO plane and individual Bi atoms can be clearly observed. Fig.1C presents the averaged scanning tunneling spectra (STS) of four samples, with pseudogap magnitudes $\Delta_{PG}$ ≈80 meV, 62 meV and superconducting gap $\Delta_{sc}$≈48 meV, 39 meV. They correspond to hole concentrations approximately $p$~0.05, 0.10, 0.15 and 0.19 respectively, (see ref (*33*)), as marked by stars and arrows in the phase diagram of Fig. 1A.

It has been proposed that doping-induced correlation changes might be observable directly as an asymmetry of electron tunneling currents with bias voltage (*16*). To explore the spectral-weight transfers induced intertwined and hidden order at low energy state, we initially choose the very under-doped sample ($p$~0.05) for measurement. Fig.1F shows the density of state (DOS) mapping at 16 mV ($g(\mathbf{r},16\text{meV})$), the corresponding morphology is shown in Fig.1E. A striking observation, and the central result of this study, is the clear detection of a four-leaf clover-shaped (FLC) electron pattern at low energy, exhibiting complex internal atomic-scale structure. This low energy charge state is significantly different from the previous reported FLC state at about 400meV (*34*).

This FLC electron pattern centered around a Cu site, is oriented along the unidirectional Cu-O bond with a separation of approximately $a_0$ (where $a_0$~3.8 Å corresponds to the Bi–Bi, or Cu–Cu, distance), as shown in Figs.1(G and H). Fig.1I presents the point d$I$/d$V$ spectra along the dashed yellow line in Fig.1H. Spectra obtained at the "petal" positions of the four-leaf clover sites (positions 3 and 5 in Fig.1G) exhibit a distinct peak around 16 meV (marked by black arrows in Fig. 1I), while the spectra from the central Cu site (position 4 in Fig.1G) lack this feature, retaining only an asymmetric differential conductance background near the Fermi level. This novel electronic structure and the behavior of low-energy unoccupied state may be related to crucial physics in high-temperature quasi-2D superconductors, such as the emergency of specific carriers and a new weakly bonding in-gap state or energy band near the Fermi level. This may be associated with the frustrated background induced by antiferromagnetic correlations or Coulomb effects in the parent insulating phase.

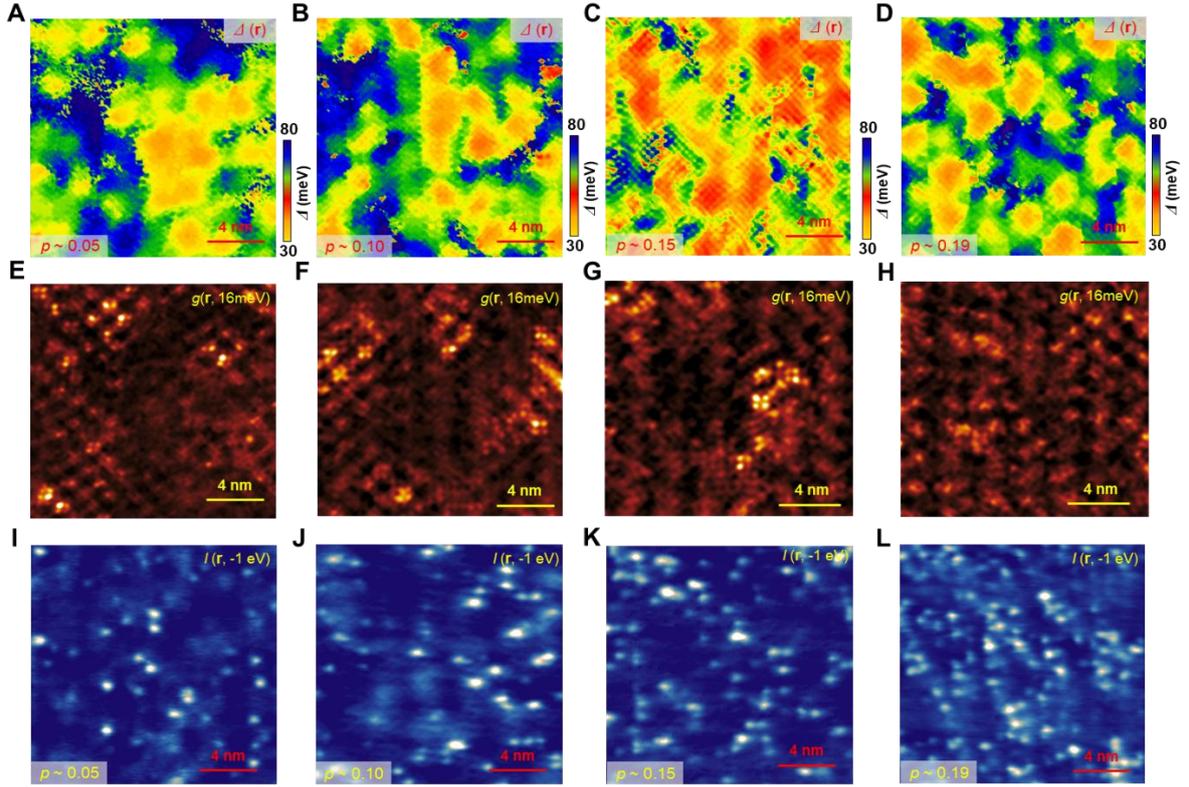

**Fig.2. Evolution of FLC charge state with increasing the holes doping.** (**A** to **D**) Superconducting gap ($\Delta(\mathbf{r})$) map of the four samples, where the local $\Delta(\mathbf{r})$ is the half energy difference between the SC coherence peaks. (**E** to **H**) Conductance map of $g(\mathbf{r},16\text{ meV})$ in the same field of view as in (A to D). The set-up conditions are listed in Methods part. (**I** to **L**) The corresponding $dI/dV$ mapping of the interstitial oxygen, which are acquired using bias voltage $V = -1$ V, tunnelling current $I = 100$ pA, respectively. All data acquired at $T = 4.2$ K.

Next, we investigate the doping dependence of this novel FLC charge order. Figs.2, A to D, show the superconducting gap ($\Delta(\mathbf{r})$) map at different doping levels, while the corresponding conductance map of $g(\mathbf{r}, 16\text{meV})$ are shown in Figs.2(E to H). Remarkably, the number of FLC electron patterns decrease sharply as hole concentration increases, and vanishes in large-scale $dI/dV$ mapping in the overdoped region (see Fig.2H). Clearly, the formation of the FLC charge order is highly sensitive to holes doping and may closely tie to the local charge or spin background. These results suggest that excessive dopants, through their intrinsic degrees, interfere with the orbital selectivity of the FLC charge order, weakening the potential landscape that facilitates FLC state formation. And, the corresponding conductance map of $g(\mathbf{r}, -16\text{meV})$ and $g(\mathbf{r}, 0\text{meV})$ are shown in **Fig.S1** and **S2**, which exclude the possibility of impurity states (*35*). Moreover, mapping the interstitial oxygen distribution (*36*) (type-B oxygen positioned closer to the BiO surface) in these four samples reveals a transition from sparse in the underdoped regime to tightly clustered in the overdoped regime (Fig.2E to H). We do not observe direct correlation between the location of FLC state and the location of doped oxygen.

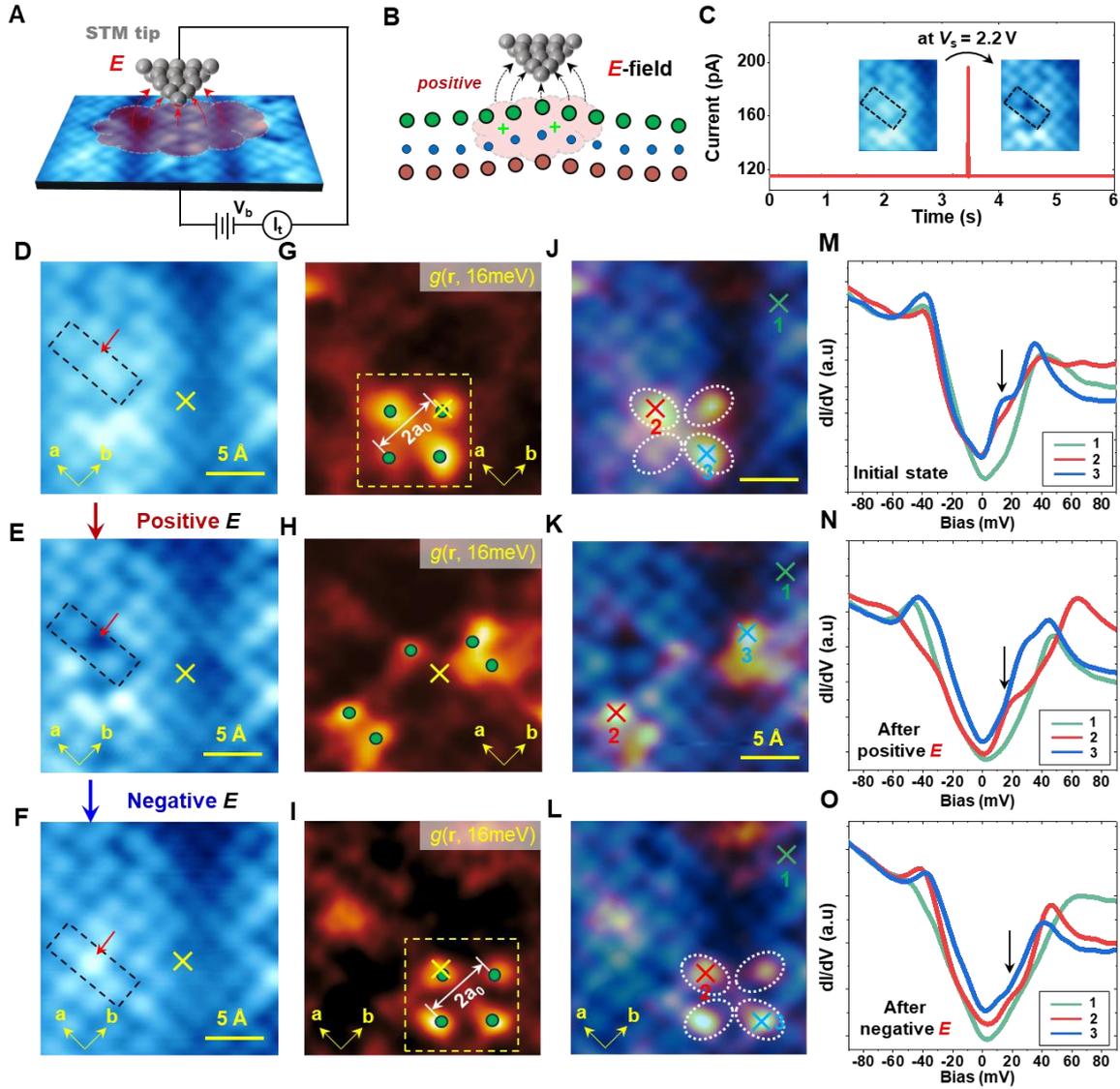

**Fig.3. Atomic manipulation of FLC charge state through local electric field of the tip.** (**A** and **B**) Illustration of the field-induced manipulation of charge carrier and surface atoms. (Bi: green; O: blue; Cu: red; charge carrier: green cross). (**C**) Current as a function of time at set point = 115 pA. Abrupt current jump (about 80 pA) corresponds to the manipulation of charge carrier and surface atoms below the tip through increasing the $V_s$. Inset: the topography images before and after manipulation. (**D** to **F**) Constant current images $I(r,E)$, ($V_{set}$ = −200 meV; $I_{set}$ = 60 pA) of reversible surface Bi atoms manipulation upon treatment. The atoms in the black dashed box are manipulated and the yellow cross represents the fixed position of the tip during electric field manipulation. (D) shows the topography before applying any electric field, while (E and F) display the topography after applying a slowly increasing positive and negative bias voltage, respectively, with a threshold voltage for the positive bias is greater than 2.2 V and for the negative bias is greater than 4.0 V. (**G** to **I**) The d$I$/d$V$ mapping $g(r, 16$ meV$)$ in the same field view of (D to F). The corresponding d$I$/d$V$ mapping $g(r, -16$ meV$)$ and $g(r, 0$ meV$)$ are shown in **Fig.S3**. The FLC charge order is highlighted by the yellow dashed box. The size of each FLC charge order is $2a_0$ along $a/b$ axis. (**J** to **L**) the overlaid constant current images on the d$I$/d$V$ mapping $g(r, 16$ meV$)$. White dashed circle represents the lobe of FLC pattern. (**M** to **O**) the d$I$/d$V$ spectra at the location marked by three crosses in (J to L), stacked along the vertical direction for better clarity and the black arrow indicate the characteristic energy scale of well-defined FLC state. All data acquired at $T$ = 4.2 K.

**Atomic manipulation of FLC charge state through local electric field of the tip.**

In contrast to previously observed relatively long charge orders (*e.g.*, stripe, checkerboard) in copper oxides (*13-26*), this novel localized isolated FLC charge state offers potential for atomic-scale manipulation. To explore the possibility of manipulating this isolated FLC charge order at the atomic scale, we applied a strong local electric field to the sample through the tip, the schematic of this method is shown in Fig.3(A to B). In constant current mode ($I_{set}$=120 pA), we maintained the tip position above a 'lobe' of a FLC pattern (marked by yellow cross in Fig.3D). When the positive bias or negative bias exceeded the certain threshold value (2.2 V and -4.0 V), a large current jump (about ~80 pA) occurred which is a symbol of completing the atomic-scale manipulation (see Fig.3D).

Figure 3(D to F) show the constant current images $I(r,E)$, ($V_{set} = -200$ meV; $I_{set} = 60$ pA) of initial state (Fig.3D), after positive $E$ (Fig.3E) and negative $E$ manipulation (Fig.3F), respectively. Significant atomic changes were observed in the vicinity of the tip, indicated by dashed black boxes in each image. The corresponding d$I$/d$V$ mapping ($g(r, 16$ meV$)$) are shown in Fig.3G to H. After performing d$I$/d$V$ mapping over the same field of view (Fig.3D to F), the FLC pattern was no longer observed and it seemed to have been replaced by a disordered electronic state (see Fig.3G and H). The large current variation, likely induced by the strong electric field of the tip, suggests that the electric field facilitates the overcoming of potential barriers driven by charge or spin background-induced frustration, allowing the localized specific electronic states near the tip to transition dramatically.

After applying a negative bias of approximately 4.0V at the same position, the topography almost restored (as shown in Fig.3F), and a new FLC pattern reappeared in the d$I$/d$V$ ($r$, 16 meV), with only a spatial shift compared to the position of FLC pattern before manipulation. Throughout the process, the tip remained highly stable. Fig.3M to O show three representative d$I$/d$V$ spectra, with crosses and numbers indicating the locations where the spectra were taken (Fig.3J to L). In the well-defined FLC charge state, the d$I$/d$V$ spectra similar to that shown in Fig.1I, with a distinct low-energy spectral feature (marked by the black arrow). After the splitting induced by positive electric-field effect, these electronic patterns no longer display FLC charge order characteristics; however, at certain high-intensity DOS locations, spectra similar to those of the FLC charge order are still observed.

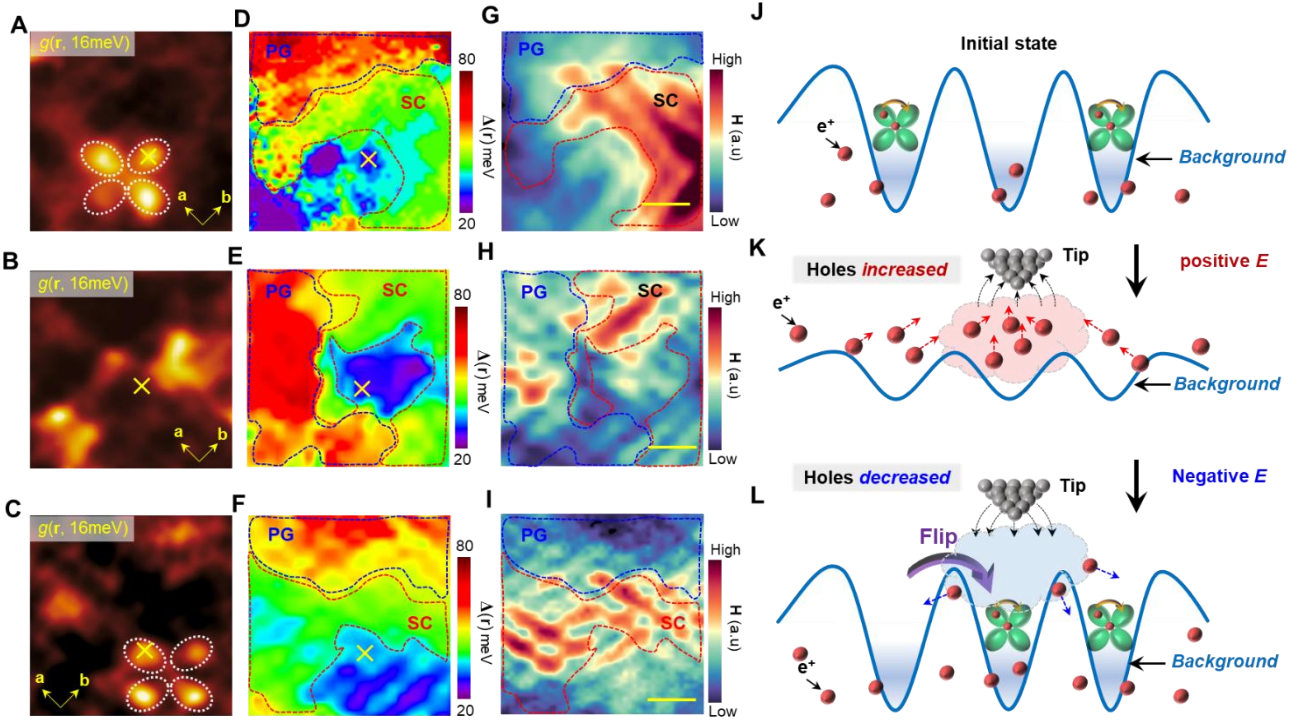

**Fig.4. | Tip-electric field manipulation of atom-site-dependent gap and superfluid density variations, and the toy model.** (**A** to **C**) The d$I$/d$V$ mapping $g(r, 16$ meV$)$ as the same FOV of Fig.3D to I. The yellow cross represents the fixed position of the tip during electric field manipulation. (**D** to **F**) Gap mapping of the same FOV. The nanoscale pseudogap region and superconducting region are enclosed by dashed boxes, respectively. The corresponding characteristic d$I$/d$V$ spectra are shown in Fig.S4. (**G** to **I**) The superfluid mapping of the same FOV by extract relative coherent-peak height. (**J** to **L**) Toy model of the variation in hole concentration induced by tip-electric field manipulation leads to the disappearance and reappearance of the FLC state, as well as its spatial displacement. Scale bar, 5Å. All data acquired at $T$ = 4.2 K.

We have successfully achieved atomic-scale manipulation of isolated FLC charge states. To further investigate the relationship between FLC charge states and superconductivity, as well as the underlying physics behind the manipulation, we acquired the corresponding spatially resolved mapping of the superconducting energy gap (see Fig.4D to F) and superfluid density (see Fig.4G to I). Fig.4D to F, reveal significant inhomogeneity in the energy gap near the FLC state, demonstrating that we not only manipulated the FLC charge states but also manipulated superconductivity at the atomic scale. We observed a dramatic change in the superconducting gap, Δ(**r**) and superfluid density (height of the coherent peak), before and after tip-induced electric field manipulation. This suggests that the tip controls the opening of new conduction pathways for charge carriers near the FLC charge order region.

**Discussion and outlook**

Based on the Mott insulator background (*2*), we propose a simplified model: at the very low doping level, the strong correlation effects lead to electron localization, and local Coulomb screening potential is weak (*37*). This results in the formation of a potential well, which easily traps the FLC charge state, as illustrated in Fig.4J. After applying the positive electric field, the local chemical environment is dramatically changed, disrupting the potential well and thus hindering the formation of the FLC electron pattern (as shown in Fig. 4B). Remarkably, this local disruption of the FLC pattern aligns with our earlier findings on overdoped samples with $p$~0.19 (see Fig.2H). This indicates that a positive electric field between the sample and the tip creates a nanoscale hole-enriched region (as illustrated in Fig.4K), which significantly increases the local doping level. The enhanced screening effect weakens the background pinning effect, leading to the disappearance of the FLC charge order. Conversely, applying a negative electric field results in a hole-depleted region, forming an underdoped background. This strengthens the pinning effect from the residual antiferromagnetic potential well, making it more likely for holes to be localized at specific sites and forming FLC charge states (as shown in Fig. 4L).

These FLC charge order do not compete with the universally observed stripe and checkerboard patterns, which have a size of approximately $4a_0$ and are considered to be the localized Cooper pairs lacking long-range coherence (*33,35,38,39*). Instead, the FLC patterns often appear as isolated $2a_0$ islands outside other ordered phases, possibly representing the smallest real-space electronic pattern of local Cooper pairs in underdoped cuprates (*40*).

These relatively isolated FLC charge order resemble an individual electronic cluster glass, with a more pronounced low-energy spectral weight compared to the stripy patterns. Previously, charge-ordered states were often associated with in-gap states in the pseudogap region (*6*), here we propose that these FLC charge order dominate low-energy carrier excitations, with an energy scale smaller than the in-gap states, on the order of tens of meV. This could also represent a preliminary stage of charge-order state, which, due to the doping level, inhomogeneity, and other influence of the chemical environment, has not yet fully developed into a specific long-range ordered charge phase. By forming the FLC electron pattern near the Cu spins, the doped holes achieve an optimized situation that minimizes strong onsite Coulomb repulsion for motion, while still benefiting from the antiferromagnetic superexchange interaction for pairing. As doping further increases, densely packed and stacked checkerboard and ladder patterns dominated by higher-energy excitations correlated to Bogoliubov-particle (*8*) emerge, along with global phase coherence and superconductivity. Thus, these $2a_0$-sized FLCs charge order are likely the smallest precursor units for the formation of incoherent Cooper pairs and the initial form of the ordered quantum electronic-liquid phase. Our work provides fresh insight into the microscopic origin of electronic inhomogeneity and charge-ordered state in high-$T_c$ cuprates.

**Acknowledgments**

We thank F.C. Zhang, K. Jiang and W.Q. Chen for help discussions. We thank Genda Gu provide the Bi-2212 single crystals.

**Funding:** This work was financially supported by National Key R&D Program of China (2024YFA1408102), the National Basic Research Program of China (Grant No. 2022YFA1403101), the National Natural Science Foundation of China (Grant No. 12141402, 12404156), the National Key Research and Development Program of China (Grant No.2022YFA1403100), the China Postdoctoral Science Foundation under Grant Number: BX20240151 and 2024M761277), and the Shenzhen Basic Research Project (Natural Science Foundation) (Grant No. JCYJ20240813100513018).

**Author contributions:** Q-K.X., Y.J.S., and Z.C.W. initiated the idea and designed the studies. Z.C.W., and F.Y.Y. performed the STM/STS experiments. Z.C.W. analysis the data with the help of F.Y.Y., Y.C.H, Z.X.W., Z.Y.S., M.Q.R., and Z.Y.C.. Z.C.W. and F.Y.Y. perform the oxygen annealing of Bi-2212 samples. All authors contributed to the scientific discussions. Z.C.W., F.Y.Y., Y.C.H., Z.X.W., J.F.J., Y.J.S. and Q-K.X. wrote the paper with contributions from all authors.

**Competing interests:** The authors declare that they have no competing interests.

**Data and materials availability:** All data needed to evaluate the conclusions in the paper are present in the paper and/or the Supplementary information.

**Supplementary Materials:** Materials and Methods; Figs. S1 to S4